\documentstyle[aps,amsfonts,twocolumn,prl,floats,epsfig]{revtex}

\begin{document}


\title{Trapping oscillations, discrete particle effects
       and kinetic theory of collisionless plasma}

\author{F.~Doveil$^a$\cite{bylineFD}\cite{bylineEmail},
        M-C.~Firpo$^a$\cite{bylineEmail},
        Y.~Elskens$^a$\cite{bylineEmail},
        D.~Guyomarc'h$^a$,
        M.~Poleni$^b$ and
        P.~Bertrand$^b$\cite{bylineEmail}
        }
\address{$^a$Equipe turbulence plasma,
                    Physique des interactions ioniques et
                    mol{\'e}culaires, \\
                    Unit{\'e} 6633 CNRS--Universit{\'e} de Provence, \\
                    case 321, Centre de Saint-J{\'e}r{\^o}me,
                    F-13397 Marseille cedex 20
         \\
         $^b$Laboratoire de physique des milieux ionis{\'e}s
                et applications, \\
                Unit{\'e} 7040 CNRS--Universit{\'e} H. Poincar{\'e}, Nancy I, \\
                BP 239, F-54506 Vand{\oe}uvre cedex, France}
\date{preprint TP99.11}
\maketitle

\begin{abstract}
Effects induced by the finite number $N$ of particles on the
evolution of a monochromatic electrostatic perturbation in a
collisionless plasma are investigated. For growth as well as
damping of a single wave, discrete particle numerical simulations
show a $N$-dependent long time behavior which differs from the
numerical errors incurred by vlasovian approaches and follows from
the pulsating separatrix crossing dynamics of individual
particles.
\\%
kinetic theory  \\%
wave-particle interaction  \\%
self-consistent field  \\%
particle motions and dynamics \par \noindent%
52.35.Fp (Plasma: electrostatic waves and oscillations) \\%
52.65.-y (Plasma simulation)\\%
52.25.Dg (Plasma kinetic equations) \\%
\end{abstract}


\section{Introduction}

It is tempting to expect that kinetic equations and their
numerical simulation provide a fair description of the time
evolution of systems with long-range or `global' interactions. A
typical, fundamental example is offered by wave-particle
interactions, which play a central role in plasmas. In this Letter
we test this opinion explicitly.

Collisionless plasma dynamics is dominated by collective
processes. Langmuir waves and their familiar Landau damping and
growth \cite{Landau} are a good example of these processes, with
many applications, e.g. plasma heating in fusion devices and
laser-plasma interactions. For simplicity we focus on the
one-dimensional electrostatic case, traditionally described by the
(kinetic) coupled set of Vlasov-Poisson equations
\cite{ONeil65,warmbeam}. The current debate on the long-time
evolution of this system shows that further insight in this
fundamental process is still needed \cite{Longtime}.

The driving process (induced by the binary Coulomb interaction
between particles) is the interaction of the electrostatic waves
in the plasma with the particles at nearly resonant velocities,
which one analyses canonically by partitioning the plasma in bulk
and tail particles. Langmuir modes are the collective oscillations
of bulk particles, with slowly varying complex amplitudes in an
envelope representation~; their interaction with tail particles is
described by a self-consistent set of hamiltonian equations
\cite{Antoni}. These equations already provided an efficient basis
\cite{Drummond} for investigating the cold beam plasma instability
and exploring the nonlinear regime of the bump-on-tail instability
\cite{Doxas}. Analytically, they were used to give an intuitive
and rigorous derivation of spontaneous emission and Landau damping
of Langmuir waves \cite{Zekri}. Besides, as it eliminates the
rapid plasma oscillation scale $\omega_{\rm p}^{-1}$, this
self-consistent model offers a genuine tool to investigate
long-time dynamics.

As we follow the motion of each particle, we can also address the
influence of the finite number of particles in the long run. This
question is discarded in the kinetic Vlasov-Poisson description,
for which the finite-$N$ correction is the Balescu-Lenard equation
\cite{Spohn} formally derived from the accumulation of weak binary
collisions, with small change of particle momenta. It implies a
diffusion of momenta, driving the plasma towards equilibrium.
However, when wave-particle coupling is dominant, the
Balescu-Lenard equation is not a straightforward approach to
finite-$N$ effects on the wave evolution.

Here we investigate direct finite-$N$ effects on the
self-consistent wave-particle dynamics. It is proved \cite{Firpo}
that the kinetic limit $N \to \infty$ commutes with evolution over
arbitrary times. As one might argue that finite $N$ be analogous
to numerical discretisation in solving kinetic equations, we also
integrate the kinetic system with a `noise-free' semi-lagrangian
solver \cite{Bertrand}. In this Letter we compare finite grid
effects of the kinetic solver and granular aspects of the
$N$-particles system, whose evolution is computed with a
symplectic scheme \cite{Doxas}.

We discuss the case of one wave interacting with the particles.
Though a broad spectrum of unstable waves is generally excited
when tail particles form a warm beam, the single-wave situation
can be realized experimentally \cite{Tsunoda} and allows to leave
aside the difficult problem of mode coupling mediated by resonant
particles \cite{Laval}.

\section{Self-consistent wave-particle model and kinetic model}

Consider a one-dimensional electrostatic potential perturbation
$\Phi(z,\tau) = [\phi_k (\tau) \exp i( k z-\omega_k \tau) + {\rm
c.c.}]$ %
(where c.c. denotes complex conjugate), with complex envelope
$\phi_k$, in a plasma of length $L$ with periodic boundary
conditions (and neutralizing background). Wavenumber $k$ and
frequency $\omega_k$ satisfy a dispersion relation
$\epsilon(k,\omega_k) = 0$. The density of $N$ (quasi-)resonant
electrons is $\sigma(z,\tau) = (nL/N) \sum_{l=1}^N
\delta(z-z_l(\tau))$, where $n$ is the electron number density and
$z_l$ is the position at time $\tau$ of electron labeled $l$ (with
charge $e$ and mass $m$). Non-resonant electrons contribute only
through the dielectric function $\epsilon$, so that $\phi_k$ and
the $z_l$'s obey coupled equations \cite{NoteEquations}
\begin{eqnarray}
  d \phi_k / d \tau
  &=&
  {{i n e} \over
           {\epsilon_0 k^2 N (\partial \epsilon/ \partial \omega_k)}}
  \sum_{l=1}^N \exp [ -i k z_l + i \omega_k \tau]
  \label{phi}
  \\
  d^2 z_l / d \tau^2
  &=&
  (i e k/m) \phi_k \exp [i k z_l - i \omega_k \tau]
  + {\rm c.c.}
  \label{force}
\end{eqnarray}
where $\epsilon_0$ is the vacuum dielectric constant. %
With $\alpha^3 = n e^2 /[m \epsilon_0
(\partial \epsilon / \partial \omega_k)]$ \cite{NoteAlpha}, %
$t = \alpha \tau$,  $\dot{ } = d/dt$, $x_l= k z_l - \omega_k \tau$
and $V = (e k^2 \phi_k)/(\alpha^2 m)$, this system defines the
self-consistent dynamics (with $N+1$ degrees of freedom)
\begin{eqnarray}
  \dot V &=& i N^{-1} \sum_{l=1}^N \exp(- i x_l)
  \label{Vdot}
  \\
  \ddot x_l &=& i V \exp(i x_l) - i V^* \exp(-i x_l)
  \label{accel}
\end{eqnarray}
for the coupled evolution of electrons and wave in dimensionless
form. This system derives from hamiltonian $H({\bf x},{\bf
p},\zeta,\zeta^*)= \sum_{l=1}^N (p_l^2/2 - N^{-1/2} \zeta e^{ix_l}
- N^{-1/2} \zeta^* e^{-ix_l})$, where a star means a complex
conjugate and $\zeta = N^{1/2} V$. An efficient symplectic
integration scheme is used to study this hamiltonian numerically
\cite{Doxas}.

The system (\ref{Vdot})-(\ref{accel}) is invariant under two
continuous groups of symmetries. Invariance under time
translations implies the conservation of the energy ${\sf H} = H$.
The phase $\theta$ of $\zeta = |\zeta| e^{-i \theta}$ plays the
role of a position for the wave, and system
(\ref{Vdot})-(\ref{accel}) is also invariant under translations
$\theta' = \theta + a$, $x_l' = x_l + a$. This translation
invariance leads to the conservation of momentum ${\sf P} = \sum_l
p_l + |\zeta|^2$, where the contribution from the wave is
analogous to the Poynting vector of electromagnetic waves (which
is quadratic in the electromagnetic fields) \cite{NoteInvariant}.
Conservation of these invariants constrains the evolution of our
system, and we checked that the numerical integration preserves
them.

In the kinetic limit $N \to \infty$, electrons are distributed
with a density $f(x,p,t)$, and system (\ref{Vdot})-(\ref{accel})
yields the Vlasov-wave system
\begin{equation}
  \dot V = i \int e^{-ix} f(x,p,t) dxdp
  \label{Vdf}
\end{equation}
\begin{equation}
  \partial_t f
  + p \partial_x f
  + (iV e^{ix} - iV^* e^{-ix}) \partial_p f
  = 0
  \label{Vlasov}
\end{equation}
For initial data approaching a smooth
function $f$ as $N \to \infty$, the solutions of
(\ref{Vdot})-(\ref{accel}) converge to those of the Vlasov-wave
system over any finite time interval \cite{Firpo}. This kinetic
model is integrated numerically by a semi-lagrangian solver,
covering $(x,p)$ space with a rectangular mesh~: the function $f$
(interpolated by cubic splines) is transported along the
characteristic lines of the kinetic equation, i.e. along
trajectories of the original particles \cite{Bertrand}.

Let us first study linear instabilities. One solution of
(\ref{Vdot})-(\ref{accel}) corresponds to vanishing field $V_0=0$,
with particles evenly distributed on a finite set of beams with
given velocities. Small perturbations of this solution have
$\delta V = \delta V_0 e^{\gamma t}$, with rate $\gamma$ solving
\cite{Zekri}
\begin{equation}
  \gamma = \gamma_{\rm r} + i \gamma_{\rm i}
         = i N^{-1} \sum_{l=1}^N (\gamma + i p_l)^{-2}.
  \label{disp}
\end{equation}
For a monokinetic beam with velocity $U$, (\ref{disp}) reads
$\gamma (\gamma + i U)^2 = i$~; the most unstable solution occurs
for $U=0$ (with $\gamma = (\sqrt{3} + i) / 2$). For a warm beam
with smooth initial distribution $f(p)$ (normalized to $\int f dp
= 1$), the continuous limit of (\ref{disp}) yields $\gamma = i
\int (\gamma + i p)^{-2} f(p) dp$.
For a sufficiently broad distribution ($|f'(0)| \ll 1$), we obtain %
$|\gamma_{\rm r}| \gamma_{\rm r} = \gamma_{\rm r} \pi
f'(-\gamma_{\rm i})$, where $f'=df/dp$, and $\gamma_{\rm i}
\approx \pi \gamma_{\rm r} f''(0)$ for $|f''(0)| \ll \pi^{-1}$.
Except for the trivial solution $\gamma_{\rm r}=0$, other
solutions can only exist for a positive slope $f'(0)$. Then the
perturbation is unstable as the evolution of $\delta V$ is
controlled by the eigenvalue $\gamma$ with positive real part,
i.e. with growth rate $\gamma_{\rm r} \approx \gamma_{\rm L} = \pi
f'(0)>0$. Negative slope leads to the linear Landau damping
paradox~: the observed decay rate $\gamma_{\rm L}=\pi f'(0)<0$ is
not associated to genuine eigenvalues, but to phase mixing of
eigenmodes \cite{Zekri,FirpoTrans,Firpo99}, as a direct
consequence of the hamiltonian nature of the dynamics.

Now, this linear analysis generally fails to give the large time
behavior. This is obvious for the unstable case as non-linear
effects are no longer negligible when the wave intensity grows so
that the trapping frequency $\omega_{\rm b} (t) = \sqrt{2 |V(t)|}$
becomes of the order of the linear rate $\gamma_{\rm r}$.

We used the monokinetic case as a testbed
\cite{Firpo99,Guyomarch}. Finite-$N$ simulations show that the
unstable solution grows as predicted and saturates to a
limit-cycle-like behavior where the trapping frequency
$\omega_{\rm b} (t)$ oscillates between $1.2 \gamma_{\rm r}$ and
$2 \gamma_{\rm r}$. In this regime, some of the initially
monokinetic particles have been scattered rather uniformly over
the chaotic domain, in and around the pulsating resonance, while
others form a trapped bunch inside this resonance (away from the
separatrix) \cite{Guyomarch}. This dynamics is quite well
described by effective hamiltonians with few degrees of freedom
\cite{Firpo99,coldbeam}.

\begin{figure}
\centerline{
  \psfig{figure=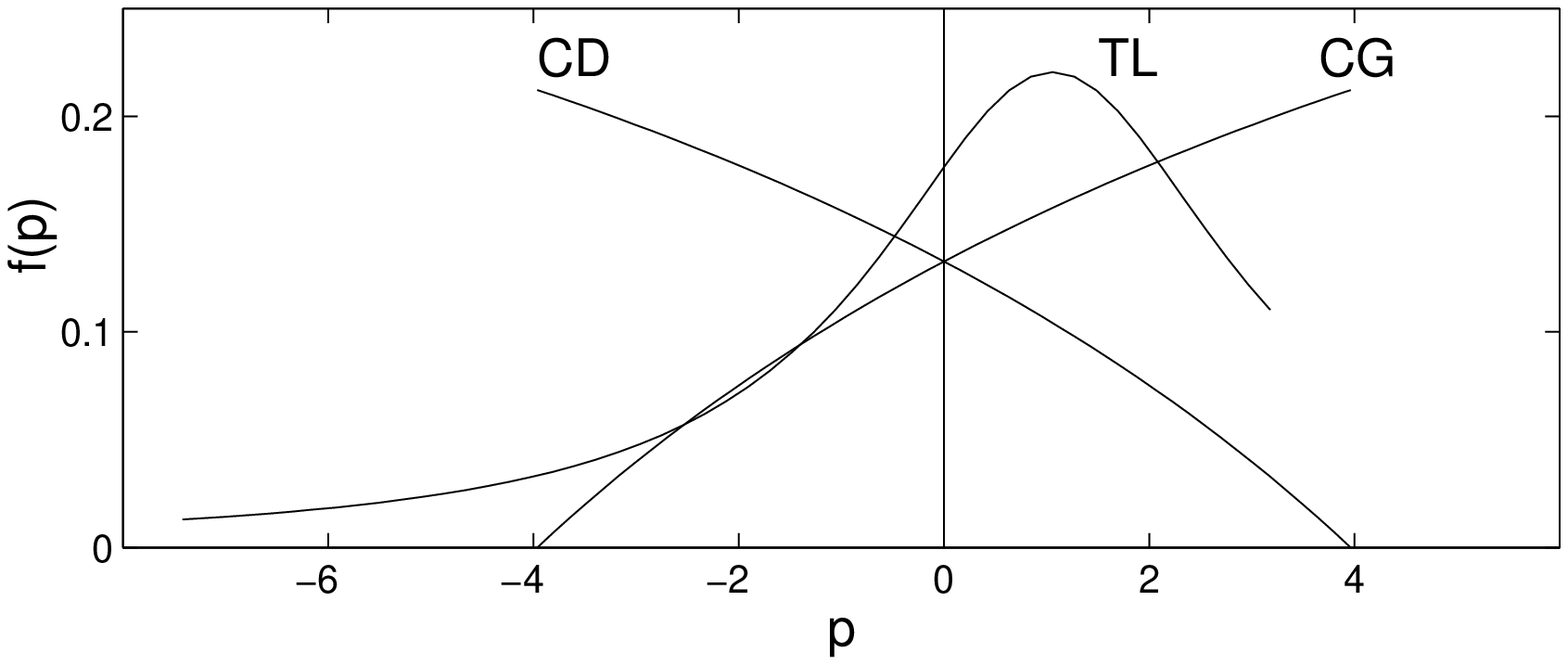,width=8cm,height=3cm}
  }
  \vskip0.2cm
\caption{ Initial velocity distributions.}
\label{fig001}
\end{figure}

In this Letter, we discuss the large time behavior of the warm
beam case, with $f'(p_0) \neq 0$ at the wave nominal velocity
$p_0=0$. Fig.~\ref{fig001} displays three distribution functions
(in dimensionless form) with similar velocity width~: {\it (i)} a
function (CD) giving the same decay rate for all phase velocities,
{\it (ii)} a function (CG) giving a constant growth rate for all
phase velocities \cite{Doxas}, {\it (iii)} a truncated Lorentzian
(TL) with positive slope $f'(0)>0$.

\section{Damping case}

For the damping case, the linear description introduces time
secularities which ultimately may break linear theory down~: the
ultimate evolution is intrinsically nonlinear, not only if the
initial field amplitude is large, as in O'Neil's seminal picture
\cite{ONeil65}, but also if one considers the evolution over time
scales of the order of the trapping time (which is large for small
initial wave amplitude). The question of the plasma wave long-time
fate is thus far from trivial \cite{Longtime}. Though some
simulations \cite{Feix} infer that nonlinear waves eventually
approach a Bernstein-Greene-Kruskal steady state \cite{BGK}
instead of Landau vanishing field, the answer should rather
strongly depend on initial conditions. Our $N$-particle, 1-wave
system is the simplest model to test these ideas.

\begin{figure}
\centerline{
  \psfig{figure=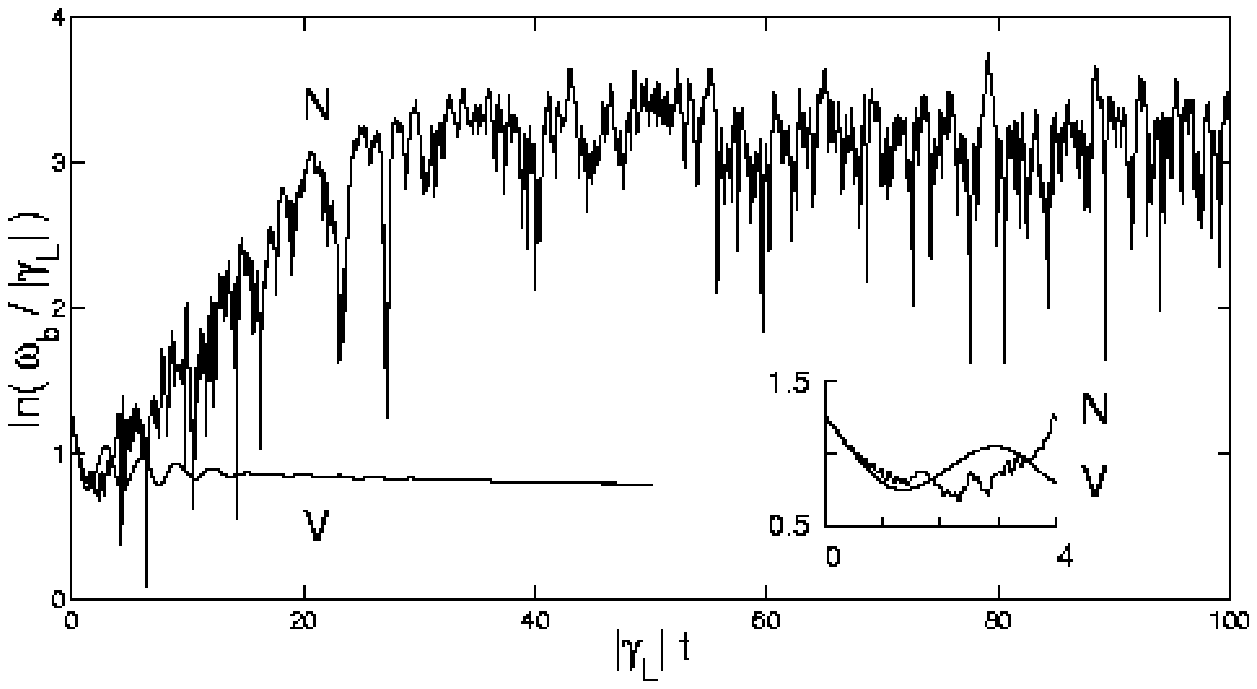,width=8cm,height=4cm}
  }
\vskip0.2cm %
\caption{ Time evolution of
 $\ln(\omega_{\rm b} (t) /|\gamma_{\rm L}|)$
 for a CD velocity distribution and
 initial wave amplitude below thermal level~: %
 (N) $N$-particles system with $N=32000$, %
 (V) kinetic scheme with $32 \times 512$ $(x,p)$ grid.
 Inset~: short-time evolution.}
\label{fig002}
\end{figure}

A thermodynamical analysis \cite{FirpoTrans} predicts that, for a
warm beam and small enough initial wave amplitude, $\omega_{\rm b}
\sim  N^{-1/2}$ at equilibrium in the limit $N\to \infty$.
Fig.~\ref{fig002} shows the evolution of a small amplitude wave
launched in the beam. The $N$-particle system (line N) and the
kinetic system (line V) initially damp the wave exponentially as
predicted by perturbation theory \cite{Zekri}, for a time of the
order of $|\gamma_{\rm L}|^{-1}$.

After that phase-mixing time, trapping induces nonlinear evolution
and both systems evolve differently. For the $N$-particle system,
the wave grows to a thermal level that scales as $N^{-1/2}$,
corresponding to a balance between damping and spontaneous
emission \cite{Zekri,FirpoTrans}. For the kinetic system, initial
Landau damping is followed by slowly damped trapping oscillations
around a mean value which also decays to zero, at a rate
decreasing for refined mesh size. Fig.~\ref{fig002} reveals that
finite-$N$ and kinetic behaviors can considerably diverge as
spontaneous emission is taken into account. The time $\tau_N$
after which the finite-$N$ effects force this divergence is found
to diverge as $N \to \infty$.

\section{Unstable case}

\begin{figure}
  \centerline{
  \psfig{figure=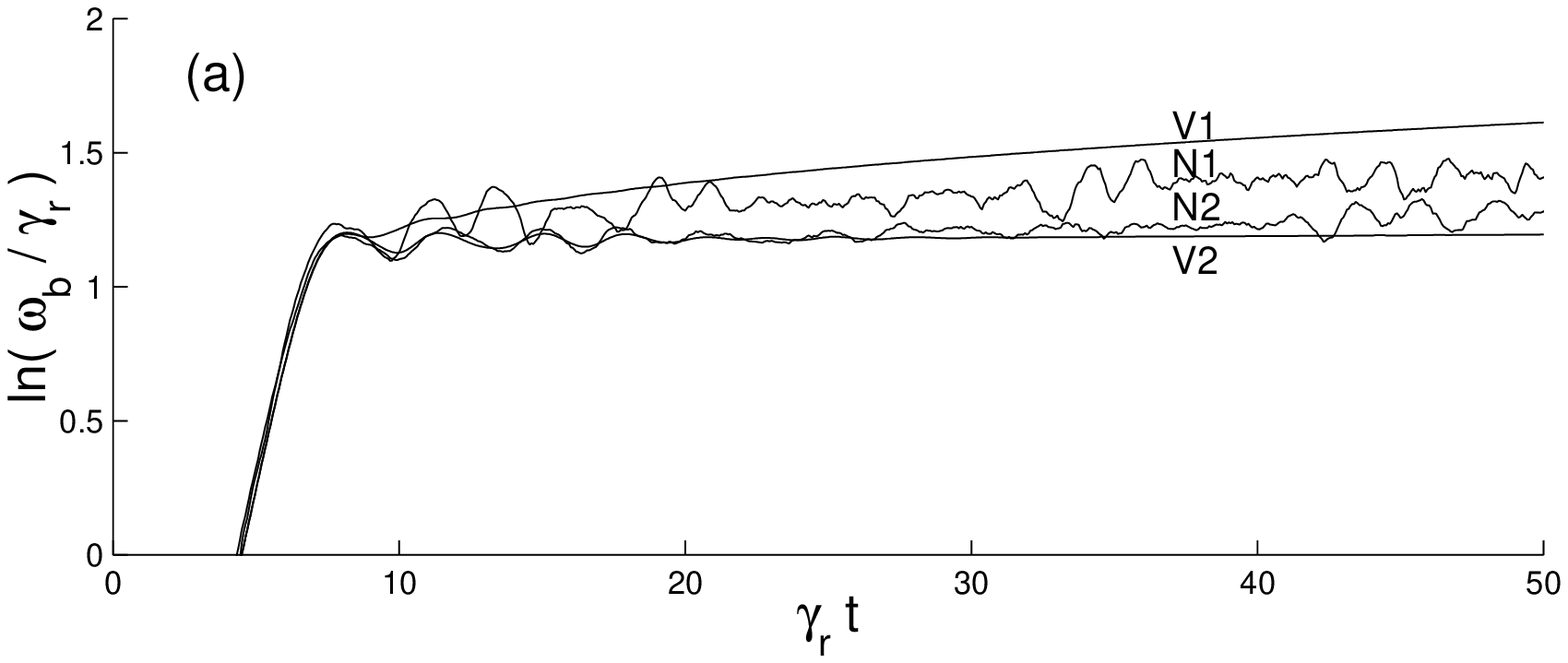,width=8cm,height=3.5cm}
  }
  \vskip1mm %
  \centerline{
  \psfig{figure=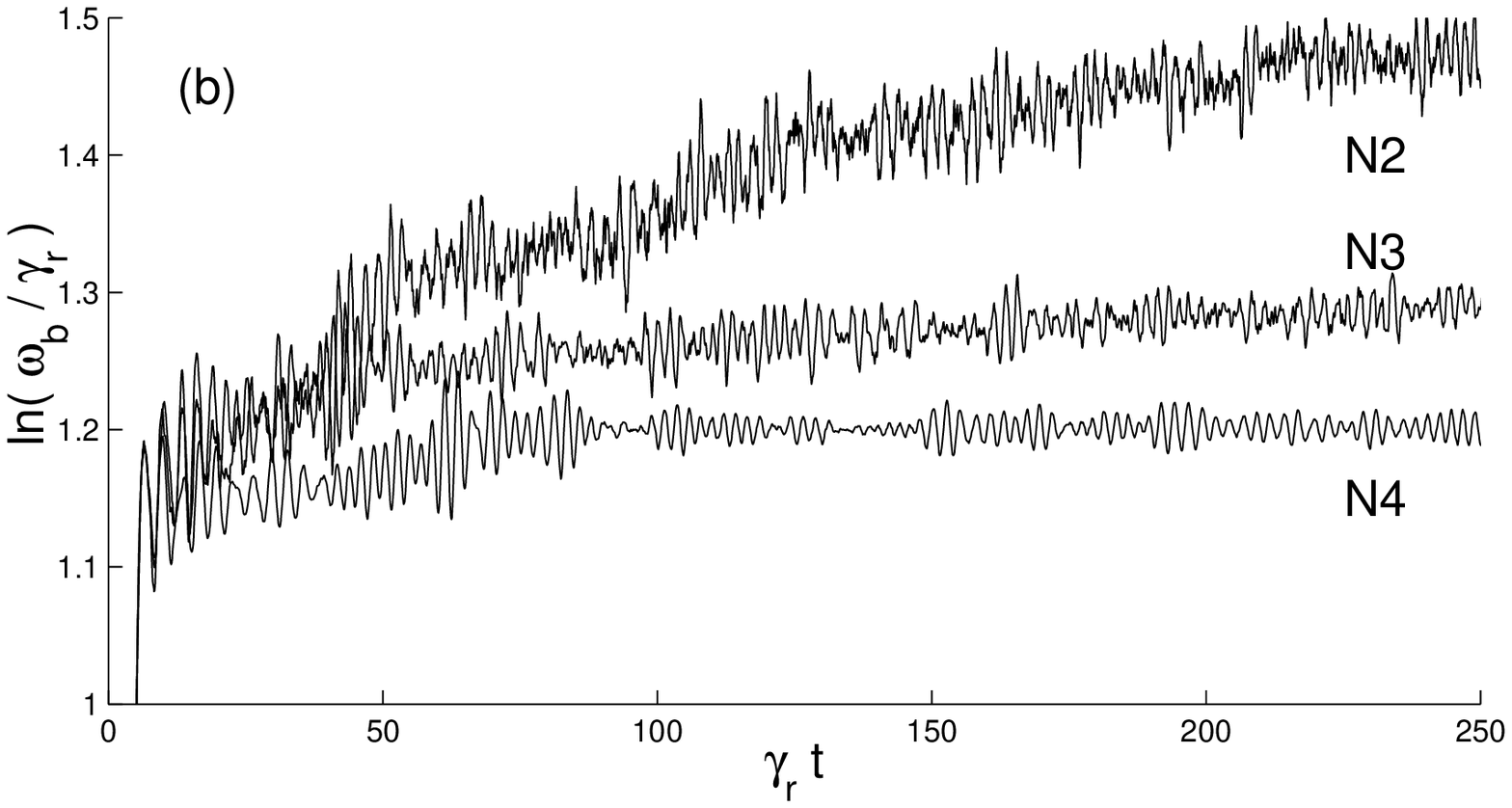,width=8cm,height=4cm}
  }
\vskip0.2cm %
\caption{ Time evolution of
 $\ln(\omega_{\rm b} (t) /\gamma_{\rm r})$.
 (a) CG initial distribution~: %
 kinetic scheme with (V1) $32 \times 128$,
 (V2) $256 \times 1024$ $(x,p)$ grid ;
 $N$-particles system with (N1) $N=128000$, (N2) $N=512000$~; %
 (b) Comparison of CG (N2) with TL initial distribution
 for (N3) $N=64000$, (N4) $N=2048000$.}
\label{fig003}
\end{figure}

Now consider an unstable warm beam ($f'(0)>0$). Line N1 (resp. N2)
of Fig.~\ref{fig003} displays $\ln(\omega_{\rm b} (t) /\gamma_{\rm
r})$ versus time for (\ref{Vdot})-(\ref{accel}) with a CG
distribution with $N=128000$ (resp. 512000) and $\gamma_{\rm r} =
0.08$. Line V1 (resp. V2) shows $\ln(\omega_{\rm b} (t)
/\gamma_{\rm r})$ versus $\gamma_{\rm r} t$ for the kinetic system
and the same initial distribution with a $32 \times 128$ (resp.
$256 \times 1024$) grid in $(x,p)$ space. All four lines exhibit
the same initial exponential growth of linear theory with less
than 1\% error on the growth rate. Saturation occurs for
$\omega_{\rm b} / \gamma_{\rm r} \approx 3.1$ \cite{warmbeam}.
Lines N1 and V1 do not superpose beyond the first trapping
oscillation after saturation. Note that, in our system,
oscillating saturation does not excite sideband Langmuir waves as
our hamiltonian incorporates only a single wave, not a spectrum.

After the first trapping oscillation, kinetic simulations exhibit
a second growth at a rate controlled by mesh size. Line V2
suggests that a kinetic approach would predict a level close to
the trapping saturation level on a time scale awarded by
reasonable integration time. This level is fairly below the
equilibrium $V_{\rm th}$ predicted by a gibbsian approach
\cite{FirpoTrans}~; such pathological relaxation properties in the
$N\to \infty $ limit seem common to mean-field long-range models
\cite{Latora}. Both kinetic simulations also exhibit a strong
damping of trapping oscillations, which disappear after a few
oscillations, whereas finite-$N$ simulations show persistent
trapping oscillations.

One could expect that finite-$N$ effects would mainly damp these
oscillations, so that the wave amplitude reaches a plateau.
Actually, we observe persistent oscillations for all $N$, and the
wave amplitude slowly grows further, whereas the velocity
distribution function flattens over wider intervals of velocity.

This spreading of particles is due to separatrix crossings, i.e.
successive trapping and detrapping by the wave \cite{Guyomarch}.
Indeed, when the wave amplitude grows (during its pulsation), it
captures particles with nearby velocity, i.e. with a relative
velocity $\Delta v_{\rm in} \approx \pm \sqrt{8|V|}$~; the trapped
particles start bouncing in the wave potential well. When the wave
amplitude decreases, particles are released, but if they
experienced only half a bouncing period, they are released with a
relative velocity (with respect to the wave) opposite to their
initial one, i.e. $\Delta v_{\rm out} \approx - \Delta v_{\rm
in}$. Now notice that a particle which has just been trapped would
oscillate at a longer period than the nominal bouncing period
(namely the one deep in the potential). Moreover, if the recently
trapped particle had just adiabatic motion in the well, it would
have to recross the separatrix when the resonance would enclose
the same area as at its trapping \cite{slowchaos}. Thus one
expects the particle to be unable to complete a full bounce, and
the fraction of particles for which $\Delta v_{\rm out} \approx -
\Delta v_{\rm in}$ is significant.

During this particle spreading process in $(x,p)$ space, the wave
pulsation is maintained by the bunch of particles which were
initially trapped, and are deep enough in the potential well to
remain trapped over a whole bouncing period. These particles form
a macroparticle, as is best seen in the case of a cold beam
\cite{coldbeam}. Note that, over long times, the macroparticle
must slowly spread in the wave resonance, following two processes.
One acts if the trapped particle motion is regular~: the trapped
motions are anisochronous, i.e. have different periods (only the
harmonic oscillator has isochronous oscillations). The other one
works if the motion is chaotic~: nearby trajectories diverge due
to chaos. Both processes contribute to the smoothing of the
particle distribution for long times, but over much longer times
than those over which we follow the system evolution and observe
the wave modulation.

This second growth after the first trapping saturation depends on
the shape of the initial distribution function. In
Fig.~\ref{fig003}(b), line N2 is the same as in
Fig.~\ref{fig003}(a), computed over a longer duration, and line N3
corresponds to $N = 64000$ with the TL distribution of
Fig.~\ref{fig001}. Although N3 corresponds to 8 times fewer
particles than N2, the final level reached at the end of the
simulation is lower. In the second growth, particles are
transported further in velocity, so that the plateau in $f(p)$
broadens with time. As the wave grows, it can trap particles with
initial velocity further away from its phase velocity. Since the
TL distribution reaches its maximum at $v \approx 1.06$ and
decreases significantly beyond this velocity (while CG is still
growing for larger $v$), fewer particles (with TL than with CG)
can give momentum to the wave when being trapped ($\sf P$ is
conserved)~; hence the second growth is slower for the TL
distribution.

We followed the evolution of the wave amplitude for N3 up to
$\gamma_{\rm r} t = 1750$~: starting from the first trapping
saturation level ($0.4 V_{\rm th}$), fluctuations persist with a
growth rate that slowly decreases as we reach $0.78 V_{\rm th}$ at
the end of the computation. Line N4 of Fig.~\ref{fig003}
corresponds to the TL distribution with 2048000 particles and
shows persistent oscillations with approximately the same
amplitude as for $N = 64000$.

\section{Conclusion}

These observations clearly indicate that the kinetic models are an
idealization and do not contain all the intricate behavior of a
discrete particles system. Now, we must also admit that the
kinetic simulation schemes do not exactly reproduce the analytic
implications of the kinetic equation. It is then legitimate to ask
whether the numerical implementation of the kinetic equations
reproduce the difference between the finite-$N$ dynamics and the
kinetic theory.

A basic property of the collisionless kinetic equation is that it
transports the distribution function $f(x,p)$ along the particle
trajectories (or characteristic lines in $(x,p)$ space). As long
as the kinetic calculation of $f$ is accurate, one expects the
kinetic scheme to follow closely the $N$-particle dynamics too.
However, the kinetic scheme is bound to depart from the analytic
predictions of the kinetic equation, because the (chaotic or
anisochronous) separation of particle trajectories implies that
constant-$f$ contours eventually evolve into complex, interleaved
shapes. This filamentation is smoothed by numerical partial
differential equation integrators, while $N$-body dynamics follows
the particles more realistically, sustaining the trapping
oscillations. Hence both types of dynamics will depart from each
other when filamentation reaches scales below the semi-lagrangian
kinetic code grid mesh.

The onset of filamentation is easily evidenced in kinetic
simulations. Indeed, whereas the kinetic equation analytically
preserves the 2-entropy $\int (1-f)f dxdp$, numerical schemes
increase entropy significantly when constant-$f$ contours form
filaments in $(x,p)$-space \cite{PoleniEntropy}. As this is also
the time at which trapping oscillations are found to damp in our
simulations, it appears that vlasovian simulations must be
considered with caution from that time on -- and it turns out that
it is also the time from which the second growth starts.

In summary, discussing the basic propagation of a single
electrostatic wave in a warm plasma, we presented 
finite-$N$ effects which do not merely result from numerical
errors and elude a kinetic simulation approach. Their
understanding depends crucially on the dynamics in phase space.
The sensitive dependence of microscopic evolution to the fine
structure of the initial particle distribution in phase space
\cite{Firpo99} implies that the interplay between limits $t \to
\infty$ and $N \to \infty$ requires some caution. Somewhat
paradoxically, refining the grid for the Vlasov simulations does
not solve this problem.

The driving process in the system evolution is separatrix
crossing, which requires a geometric approach to the system
dynamics. Further work in this direction \cite{Benisti} will also
shed new light on the foundations of common approximations, such
as replacing original dynamics (\ref{phi})-(\ref{force}) by
coupled stochastic equations, in which particles undergo noisy
transport.

\section{Acknowledgments}

The authors thank D.F. Escande for fruitful discussions, and J.R.
Cary and I. Doxas for computational assistance. MCF and DG were
supported by the French Minist{\`e}re de la Recherche. Computer
use at Institut M{\'e}diterran{\'e}en de Technologie and IDRIS was
granted by R{\'e}gion Provence-Alpes-C{\^o}te d'Azur and CNRS.
This work is part of the european network {\it Stability and
universality in classical mechanics} and CNRS GdR {\it
Syst{\`e}mes de particules charg{\'e}es} (SParCh).

%

\clearpage


\clearpage


\begin{references}
\bibitem{NoteEquations} {These equations are derived by O'Neil,
  Winfrey and Malmberg in \cite{Drummond} and, within the
  lagrangian approach, in \cite{Antoni}.}
\bibitem{NoteInvariant} {One can rewrite the hamiltonian dynamics
 (\ref{Vdot})-(\ref{accel}) using intensity-phase variables $(I,\theta)$
 for the wave, with $\zeta = \sqrt{I} e^{- i \theta}$. The total
 momentum ${\sf P} = \sum_l p_l + I$ is a linear function of the
 wave intensity and of the particle momenta, while the energy reads
 ${\sf H} = \sum_l p_l^2/2 - 2 N^{-1/2} \sum_l \sqrt{I} \cos(x_l-\theta)$.}
\end{references}
\end{document}